\documentclass[aps,PRB,twocolumn,superscriptaddress,preprintnumbers]{revtex4-2}
\usepackage[T1]{fontenc}
\usepackage{times}
\usepackage{graphicx,color}
\usepackage{amsfonts,amsmath,amssymb,amsbsy}
\usepackage[colorlinks=true, linkcolor=blue, citecolor=blue, urlcolor=blue]{hyperref}
\usepackage{float}

\begin{document}

\title{Higher-order topological quantum battery}
\author{Motohiko Ezawa}
\affiliation{Department of Applied Physics, The University of Tokyo, 7-3-1 Hongo, Tokyo
113-8656, Japan}

\begin{abstract}
A quantum battery is a next-generation energy-storage device that exploits
genuinely quantum properties such as superposition and coherence. We propose
a second-order topological quantum battery coupled to a photonic crystal
described by the breathing-Kagome model, where optical loss is present on
all lattice sites while nonlinear gains are introduced selectively at the
output sites. Under these realistic non-Hermitian conditions, the system
exhibits a clear topological phase transition manifested in both the
ergotropy and the charging power. Remarkably, perfect work extraction is
realized in the topological phase, where the maximum value of the ergotropy
is quantized and acts as a good topological-phase indicator. The triangular geometry coherently distributes energy into two quantum batteries, where mirror-symmetry-protected phase synchronization gives rise to collective energy storage.
\end{abstract}

\date{\today }
\maketitle

\textbf{Introduction: }Quantum batteries\cite{Alic,Ferra,CampRMP} constitute
a promising paradigm within quantum technologies, where energy is stored and
manipulated by exploiting genuinely quantum mechanical resources such as
superposition and coherence. These quantum features enable charging
processes that surpass their classical counterparts in terms of achievable
charging power\cite{Camp,Seah,Zhu,Ghym}, capacity\cite{Yang}, and
extractable work\cite{Shi,Ando,Monsel,Tiro}. A central figure of merit is
the ergotropy\cite{Alla,Franc} $W\left( t\right) $, which quantifies the
maximum amount of useful work that can be extracted from the quantum battery
via unitary operations. The charging power\cite{Binder,Camp,Sree} defined as 
$P\left( t\right) =dW\left( t\right) /dt$ provides a complementary measure,
characterizing the rate at which extractable energy is accumulated or
released. Quantum batteries have a broad range of potential applications,
including quantum sensing, quantum computing, and quantum communication
networks, where rapid and on-demand energy delivery is essential. Moreover,
they provide a fertile platform for exploring fundamental aspects of quantum
thermodynamics\cite{CampRMP,BinderBook}, such as work extraction,
non-equilibrium dynamics, and the role of coherence as a thermodynamic
resource.

\begin{figure}[t]
\centerline{\includegraphics[width=0.48\textwidth]{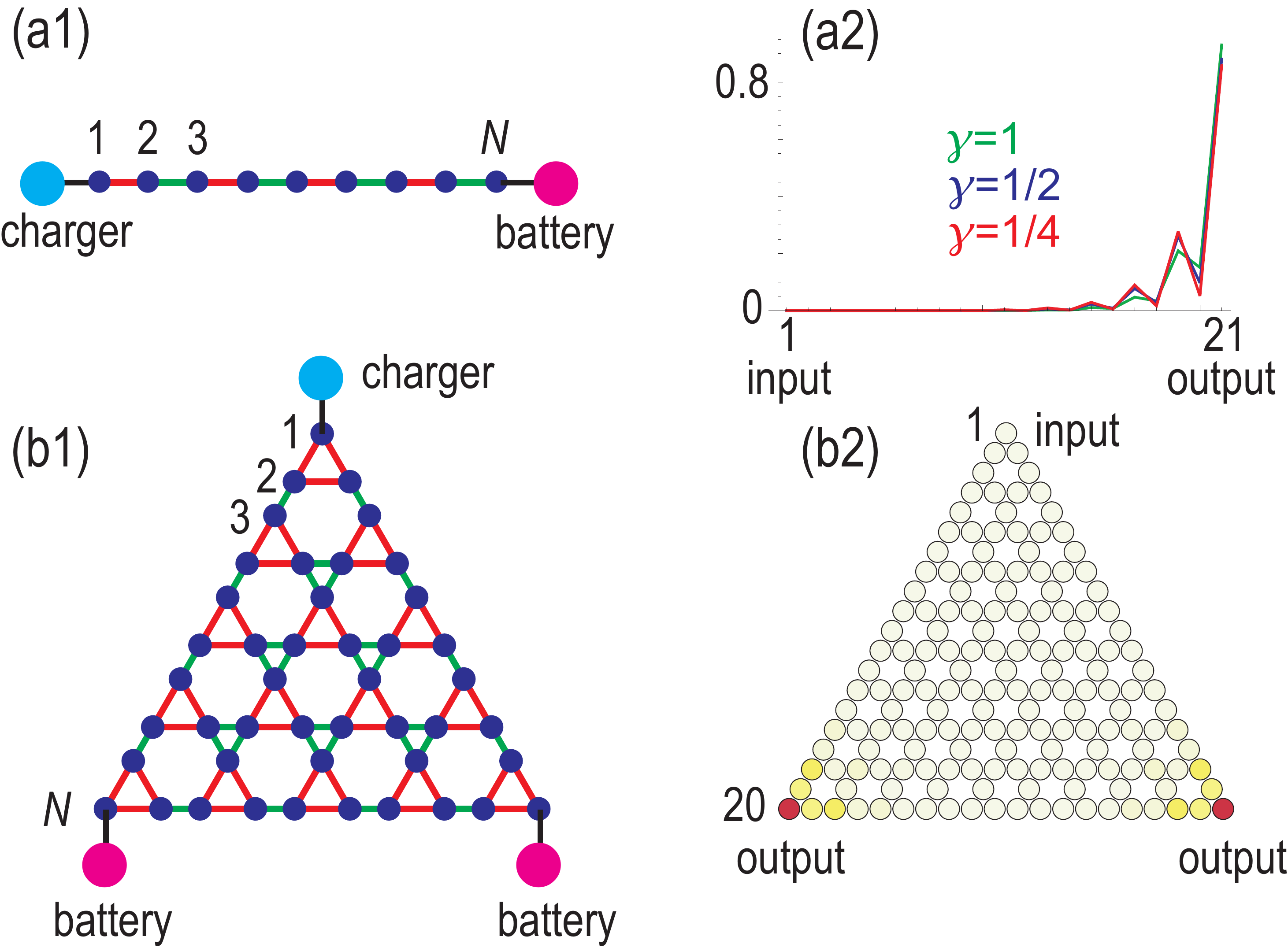}}
\caption{(a1) Illustration of a topological quantum battery based on the SSH
model. (b1) Illustration of a second-order topological quantum battery based
on the breathing-Kagome model. Red bars represent hopping $\protect\kappa %
_{A}$ and the blue bars represent hopping $\protect\kappa _{B}$. (a2) Edge
state of the topological quantum battery after relaxation. (b2) Corner
states of the second-order topological quantum battery after relaxation. We
have set $N=21$ for the SSH model and $N=20$ for the breathing-Kagome model
in numerical calculations. }
\label{FigIllust}
\end{figure}

Recently, a topological quantum battery\cite{Lu} has been proposed based on
the Su-Schrieffer-Heeger (SSH) chain, as illustrated in Fig.\ref{FigIllust}%
(a1). In the original scheme, the SSH lattice is implemented in a photonic
platform where selective loss is introduced only on one sublattice. This
concept was later generalized to a parity-time (PT) symmetric topological
quantum battery\cite{Konar,Zhou} by incorporating balanced linear gain on
the opposite sublattice, thereby exploiting non-Hermitian dynamics to
enhance charging performance. In realistic photonic lattices, however, loss
is present on all sites uniformly and optical gain is inherently nonlinear
due to saturation effects. Indeed, topological lasers\cite{Harrari,Bandres}
are realized under these realistic conditions, where non-Hermitian topology,
gain saturation, and distributed loss collectively determine the emergence
and stability of lasing edge modes. These same mechanisms may play a crucial
role in shaping the behavior of practical topological quantum batteries
realized in photonic platforms.

In this paper, we propose a second-order topological quantum battery coupled
to a photonic crystal based on the breathing-Kagome model\cite{EzawaKagome}
with the loss introduced for all sites and the nonlinear gain introduced
selectively at the output sites. Two quantum batteries are connected in a
triangle made of the breathing-Kagome lattice, as illustrated in Fig.\ref%
{FigIllust}(b1). A clear topological phase transition is realized in the
ergotropy and the charging power. Especially, the perfect work extraction is
possible in the topological phase, where the maximum value of the ergotropy
is quantized and acts as a good topological-phase indicator. Due to the
mirror symmetric configuration, two quantum batteries are charged
simultaneously and coherently.

\textbf{Model: }We consider a system made of a charger and quantum batteries
connected by a photonic crystal. The minimal and most widely studied model%
\cite{Binder,Pintos} of a quantum battery is a two-level system. The
Hamiltonian of the $j$-th quantum battery is given by\cite{Binder,Pintos}%
\begin{equation}
H_{\text{QB},j}=\frac{\hbar \omega _{\text{QB}}}{2}\sigma _{j}^{z},
\label{HamilQB}
\end{equation}%
where $\hbar \omega _{\text{QB}}$ is the energy split between the ground
state and the excited state, and $\sigma _{j}^{z}$ is the Pauli operator for
the quantum battery $j$. Typically, the quantum battery is constructed by a
superconducting qubit, a quantum dot, an atom or an NV center in diamonds.

The Hamiltonian for a photonic crystal is given by\cite{Harrari,EzawaLaser}%
\begin{equation}
H_{\text{photon}}=\sum_{n}\frac{\hbar \omega _{\text{photon}}}{2}\psi
_{n}^{\ast }\psi _{n}+\sum_{\left\langle n,m\right\rangle }M_{nm}\psi
_{n}^{\ast }\psi _{m},
\end{equation}%
where $\psi _{n}$ is a classical field representing the amplitudes of
photons at the site $n$, where $n=1,2,3,\cdots ,N$ in the system composed of 
$N$\ sites, $M_{nm}$ describes a hopping matrix between the adjacent sites $%
\left\langle n,m\right\rangle $ characterizing the photonic crystal, and $%
\omega _{\text{photon}}$ is the characteristic frequency of the photonic
crystal.

By using the rotational-wave approximation, the coupling between the quantum
battery and the photonic crystal is given by%
\begin{equation}
H_{\text{coupling}}=\sum_{j\in \text{output}}g\psi _{j}\sigma
_{j}^{+}+g^{\ast }\psi _{j}^{\ast }\sigma _{j}^{-},
\end{equation}%
where $j\in $output means that the site $j$ is connected with the quantum
battery and $g$ is the coupling constant. See Fig.\ref{FigIllust}(a1) for
the SSH model and Fig.\ref{FigIllust}(b1) for the breathing-Kagome model as
illustrations. 

The corresponding equations of motion for the quantum battery are given by%
\begin{eqnarray}
\frac{d\rho _{j}^{+}}{dt} &=&i\left( \omega _{\text{QB}}-\omega _{\text{%
photon}}\right) \rho _{j}^{+}-ig\psi _{j}^{\ast }\rho _{j}^{z},  \label{EqA}
\\
\frac{d\rho _{j}^{-}}{dt} &=&-i\left( \omega _{\text{QB}}-\omega _{\text{%
photon}}\right) \rho _{j}^{-}-ig\psi _{j}\rho _{j}^{z},  \label{EqB} \\
\frac{d\rho _{j}^{z}}{dt} &=&ig\psi _{j}\rho _{j}^{+}-g^{\ast }\psi
_{j}^{\ast }\rho _{j}^{-},  \label{EqC}
\end{eqnarray}%
while that for the photonic crystal is described by 
\begin{equation}
i\frac{d\psi _{n}}{dt}=\sum_{nm}M_{nm}\psi _{m}-i\gamma \left( 1-\xi \frac{%
\delta _{n,\text{output}}}{1+\left\vert \psi _{n}\right\vert ^{2}/\eta }%
\right) \psi _{n}-ig^{\ast }\sigma ,  \label{EqPsi}
\end{equation}%
where $\gamma $ represents the loss; $\gamma \xi $ represents the amplitude
of the optical gain via stimulated emission; $\eta $ represents the
nonlinearity; $\delta _{n,\text{output}}$ stands for the optical pump
introduced at the output site. The system turns into the linear model in the
limit $\eta \rightarrow \infty $. The system turns into a Hermitian model
for $\gamma =0$. We call the term\ proportional to $\gamma $ the loss term
and the term\ proportional to $\gamma \xi $\ the nonlinear gain term\cite%
{Harrari,EzawaLaser}. The gain is introduced by optical pumping. We
introduce the gain only for the output sites $j$. On the other hand, the
loss is uniformly introduced for the all sites, which is contrasted in the
original work, where the loss is introduced only for one sublattice. We set $%
\omega _{\text{QB}}=\omega _{\text{photon}}$\cite{Ob,Ahu}, which is
experimentally feasible\cite{Mang,ZXu,WLee}. We prepare the initial
conditions, where $\psi _{1}=1$ and otherwise $\psi _{n}=0$ with respect to
the photonic crystal in Fig.\ref{FigIllust}(a1) and (b1), while the quantum
battery is the ground state.

\begin{figure}[t]
\centerline{\includegraphics[width=0.48\textwidth]{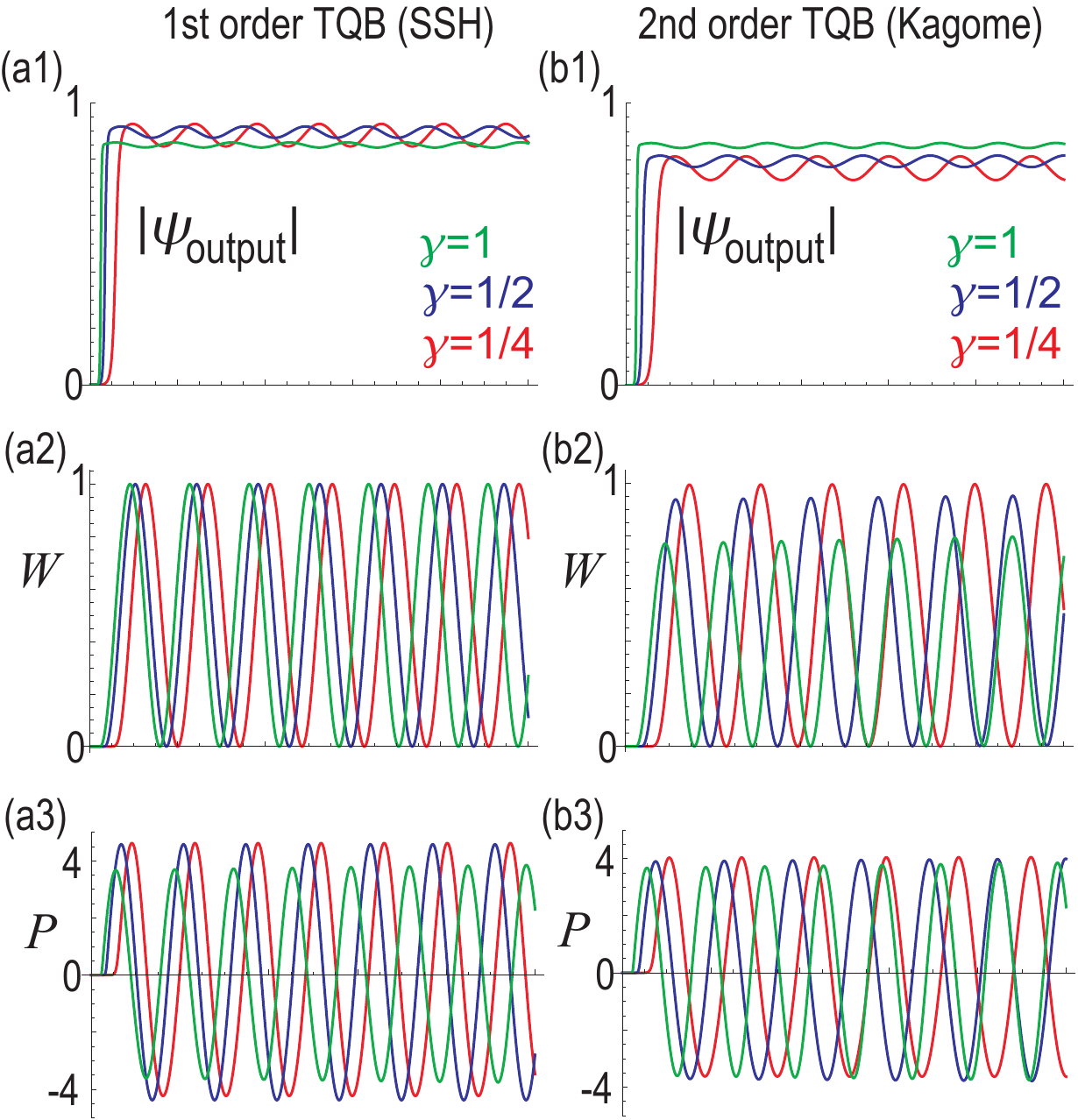}}
\caption{(a1) Temporal evolution of the amplitude at the edge site of the
topological quantum battery. (b1) Temporal evolution of the amplitude at the
corner sites of the second-order topological quantum battery. (a2) Temporal
evolution of the ergotropy $W(t)$ of the topological quantum battery. (b2)
Temporal evolution of the ergotropy of the second-order topological quantum
battery. The vertical axis is the $W\left( t\right) /\frac{\hbar \protect%
\omega _{\text{QB}}}{2}$. (a3) Temporal evolution of the charging power $%
P(t)=dW(t)/dt$ of the topological quantum battery. (b3) Temporal evolution
of the charging power $P(t)$ of the second-order topological quantum
battery. The vertical axis is the charging power $P\left( t\right) /\frac{%
\hbar \protect\omega _{\text{QB}}}{2\tau_0}$. We \ have set $\protect\kappa =\varepsilon_0$, $%
g=0.1\varepsilon_0$, $\protect\eta =1$\ and $\protect\xi =2$, where $\tau_0$ is the time unit and $\varepsilon_0$ is the energy unit.}
\label{FigDynamics}
\end{figure}

\textbf{SSH model}: A topological quantum battery was originally proposed
based on the SSH model by using the resolvent method for the bulk system,
where the loss is introduced selectively only at one sublattice and gain is
not introduced. Here, we study the SSH model under different conditions,
where the loss is introduced for all sites and the nonlinear gain is
introduced selectively at the output site. We note that a topological laser
based on the SSH model has been studied\cite%
{Schome,Weimann,Jean,Ota18,Parto,Zhao,Malzard,MalzardOpt,EzawaLaser} under
these conditions. We show the band structure for a finite chain in Fig.\ref%
{FigMax}(a2). We set the number of sites to be odd as shown in Fig.\ref%
{FigIllust}(a1), where there is only one edge state in the topological phase.

The equations of motion are derived by setting $M_{nm}$\ to be the hopping
matrix describing the SSH model. It follows from Eq.(\ref{EqPsi}) that the
explicit equations for a finite chain with length $N$\ are given by%
\begin{eqnarray}
i\frac{d\psi _{2n-1}}{dt} &=&\kappa _{A}\left( \psi _{2n}-\psi
_{2n-1}\right) +\kappa _{B}\left( \psi _{2n-2}-\psi _{2n-1}\right)   \notag
\\
&&-i\gamma \left( 1-\xi \frac{\delta _{n,\text{output}}}{1+\left\vert \psi
_{2n-1}\right\vert ^{2}/\eta }\right) \psi _{2n-1},  \label{SSH1} \\
i\frac{d\psi _{2n}}{dt} &=&\kappa _{B}\left( \psi _{2n+1}-\psi _{2n}\right)
+\kappa _{A}\left( \psi _{2n-1}-\psi _{2n}\right)   \notag \\
&&-i\gamma \left( 1-\xi \frac{\delta _{n,\text{output}}}{1+\left\vert \psi
_{2n}\right\vert ^{2}/\eta }\right) \psi _{2n}.  \label{SSH2}
\end{eqnarray}%
It is convenient to introduce the coupling strength $\kappa $ and the
dimerization parameter $\lambda $ by 
\begin{equation}
\kappa _{A}=\kappa \left( 1+\lambda \right) ,\quad \kappa _{B}=\kappa \left(
1-\lambda \right)   \label{DimerParam}
\end{equation}%
with $|\lambda |\leq 1$. It is known that the system is topological for $%
\lambda <0$, where there is an edge state at the right-most edge site for
odd number of sites. We show the spatial profile after relaxation in Fig.\ref%
{FigIllust}(a2), where the topological edge state is formed at the right
hand side even in the presence of the quantum battery. By numerically
solving Eqs.(\ref{SSH1}) and (\ref{SSH2}), we show the dynamics of $\psi _{%
\text{output}}$ in Fig.\ref{FigDynamics}(a1).

\textbf{Breathing Kagome model}: Higher-order topological insulators are
generalization of topological insulators\cite%
{Science,Lang,Song,Bena,Schin,EzawaKagome}, where topological corner states
emerge. A typical example is realized in the breathing-Kagome model\cite%
{EzawaKagome}, which is a second-order topological insulator. We illustrate
the breathing-Kagome lattice in Fig.\ref{FigIllust}(b1).

The hopping matrix is explicitly given by\cite{EzawaKagome} 
\begin{equation}
M\left( \mathbf{k}\right) =-\left( 
\begin{array}{ccc}
0 & h_{12} & h_{13} \\ 
h_{12}^{\ast } & 0 & h_{23} \\ 
h_{13}^{\ast } & h_{23}^{\ast } & 0%
\end{array}%
\right)  \label{H3}
\end{equation}%
with 
\begin{align}
h_{12}& =\kappa _{A}e^{i\left( k_{x}/2+\sqrt{3}k_{y}/2\right) }+\kappa
_{B}e^{-i\left( k_{x}/2+\sqrt{3}k_{y}/2\right) }, \\
h_{23}& =\kappa _{A}e^{i\left( k_{x}/2-\sqrt{3}k_{y}/2\right) }+\kappa
_{B}e^{i\left( -k_{x}/2+\sqrt{3}k_{y}/2\right) }, \\
h_{13}& =\kappa _{A}e^{ik_{x}}+\kappa _{B}e^{-ik_{x}}
\end{align}%
in the momentum space, where we have introduced two hopping parameters $%
\kappa _{A}$ and $\kappa _{B}$ corresponding to upward and downward
triangles in Fig.\ref{FigIllust}(b). The equations of motions follow
explicitly from Eq.(\ref{EqPsi}). It is known that the system is topological
for $\lambda <-1/3$, where the corner states emerge as shown in Fig.\ref%
{FigMax}(b2). On the other hand, it is metallic for $-1/3<\lambda <1/3$\ and
trivial for $1/3<\lambda <1$, where there is no corner state. We show the
dynamics of $\psi _{\text{output}}$ in Fig.\ref{FigDynamics}(b1).

We show the spatial profile after relaxation in Fig.\ref{FigIllust}(b2),
where two bottom corner states are enhanced. We note that the top corner
state disappears due to the loss term. 
\begin{figure}[t]
\centerline{\includegraphics[width=0.48\textwidth]{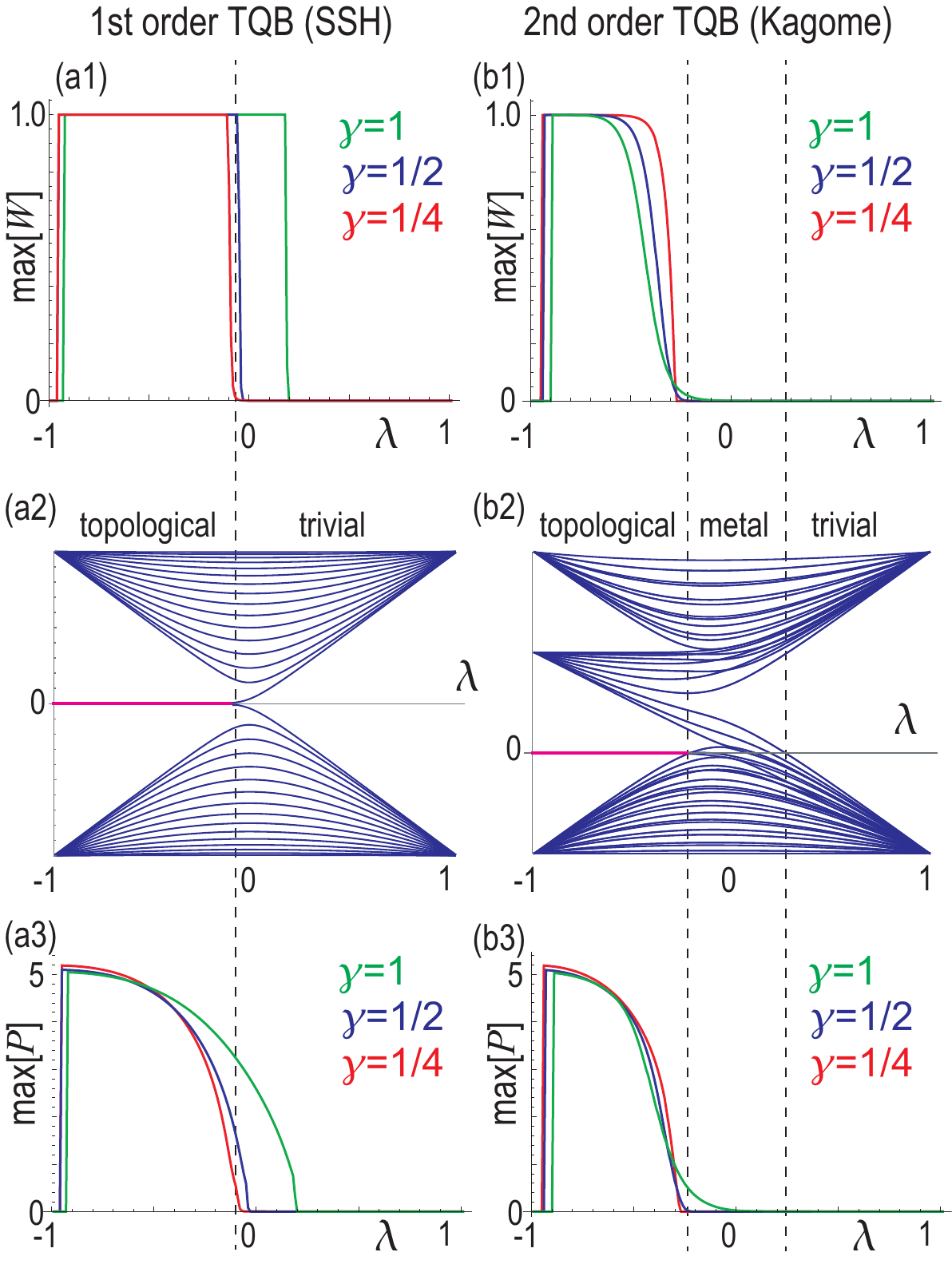}}
\caption{(a1) Maximum value of the ergotropy in the topological quantum
battery. (b1) Maximum value of the ergotropy in the second-order topological
quantum battery. The vertical axis is the Max[$W$]$/\frac{\hbar \protect%
\omega _{\text{QB}}}{2}$. (a2) Energy spectrum in the SSH model made of a
finite chain with topological edge states in red for $\protect\lambda <0$.
(b2) Energy spectrum of the breathing-Kagome model in triangle geometry with
topological corner states in red for $\protect\lambda <-1/3$. (a3) Maximum
value of the charging power in the topological quantum battery. (b3) Maximum
value of the charging power in the second-order topological quantum battery.
The vertical axis is the charging power $P\left( t\right) /\frac{\hbar 
\protect\omega _{\text{QB}}}{2\tau_0}$. The horizontal axis is the dimerization
parameter $\protect\lambda $ for (a1), (a2) and (a3). The horizontal axis is
the trimerization parameter $\protect\lambda $ for (b1), (b2) and (b3).}
\label{FigMax}
\end{figure}

\textbf{Ergotropy}: The ergotropy\cite{Alla,Franc} is a fundamental
indicator measuring the performance of the quantum battery. It is the
maximum amount of work that can be extracted from a quantum system. It is
defined by 
\begin{equation}
W_{j}\left( t\right) \equiv W_{\text{QB},j}\left( t\right) -W_{\text{QB},j}^{%
\text{passive}},  \label{Ergo}
\end{equation}%
where $W_{\text{QB},j}$ is the energy for the quantum battery and $W_{\text{%
QB},j}^{\text{passive}}$ is the energy for the passive state. They are
defined as 
\begin{eqnarray}
W_{\text{QB},j} &\equiv &\text{Tr}\left[ \rho _{\text{QB},j}\left( t\right)
H_{\text{QB},j}\right] ,  \label{WQB} \\
W_{\text{QB},j}^{\text{passive}} &\equiv &\text{Tr}\left[ \rho _{\text{QB}%
,j}^{\text{passive}}H_{\text{QB},j}\right]   \label{Wpass}
\end{eqnarray}%
where $H_{\text{QB},j}$ is the Hamiltonian (\ref{HamilQB}) for the quantum
battery, while $\rho _{\text{QB},j}\left( t\right) $ is the density matrix
for the quantum battery, and $\rho _{\text{QB},j}^{\text{passive}}$ is that
for the passive state. The passive state is such a state that has the
minimum energy by making an arbitrary unitary transformation for $\rho _{%
\text{QB},j}$ by fixing $H_{\text{QB},j}$ in Eq.(\ref{WQB}).

The passive state is constructed explicitly as follows\cite{Pusz}. We expand
the density matrix for the quantum battery by using the Pauli matrices as%
\begin{equation}
\rho _{\text{QB},j}\left( t\right) =\frac{I+\sum_{\mu =x,y,z}\rho _{j}^{\mu
}\left( t\right) \sigma ^{j}}{2},
\end{equation}%
where $\rho _{j}^{\mu }\left( t\right) $\ is a scalar. Then, Eq.(\ref{WQB})
reads 
\begin{equation}
W_{\text{QB},j}\left( t\right) =\frac{\hbar \omega _{\text{QB}}}{2}\rho
_{j}^{z}\left( t\right) .
\end{equation}%
The eigenvalues of the density matrix is given by $r_{j}^{\pm }=\left( 1\pm
r_{j}\right) /2$ with $r_{j}=\sqrt{\sum_{\mu =x,y,z}\left( \rho _{j}^{\mu
}\right) ^{2}}$. The passive state is constructed by assigning the larger
eigenvalue $r^{+}$ to the lower-energy state $\left\vert 0\right\rangle $
and the smaller eigenvalue $r^{-}$ to the higher-energy state $\left\vert
1\right\rangle $, which is%
\begin{equation}
\rho _{\text{QB},j}^{\text{passive}}=r_{j}^{+}\left\vert 0\right\rangle
\left\langle 0\right\vert +r_{j}^{-}\left\vert 1\right\rangle \left\langle
1\right\vert .
\end{equation}%
Then, Eq.(\ref{Wpass}) is calculated as%
\begin{equation}
W_{\text{QB},j}^{\text{passive}}=\frac{\hbar \omega _{\text{QB}}}{2}\left(
r_{j}^{-}-r_{j}^{+}\right) =-\frac{\hbar \omega _{\text{QB}}}{2}r_{j}.
\end{equation}%
It gives the ergotropy (\ref{Ergo}) as%
\begin{equation}
W_{j}\left( t\right) =\frac{\hbar \omega _{\text{QB}}}{2}\left( \rho
_{j}^{z}\left( t\right) +r_{j}\right) .
\end{equation}%
We have $r_{j}=1$ for the unitary evolution of the quantum battery. We
obtain $\psi _{n}$ and $\rho _{j}^{z}$ by numerically solving Eqs.(\ref{EqA}%
), (\ref{EqB}), (\ref{EqC}) and (\ref{EqPsi}). We have $W_{1}\left( t\right)
=W_{2}\left( t\right) \equiv W\left( t\right) $ due to the mirror symmetry.

We show the dynamics of the ergotropy $W\left( t\right) $ in Fig.\ref%
{FigDynamics}(a2) for the topological quantum battery and Fig.\ref%
{FigDynamics}(b2) for the second-order topological quantum battery. They
reach saturated oscillations, where the maximum value of the ergotropy is a
marker of the quantum battery. The phases and amplitudes of the two quantum
batteries are exactly identical due to the mirror symmetry of the
configuration. The maximum values of the ergotropy max[$W$] after relaxation
are shown as a function of $\lambda $ in Fig.\ref{FigMax}(a1) and (b1). We
find that they are zero for the trivial phase and nonzero for the
topological phase. Especially, it is quantized sharply for the topological
quantum battery as shown in Fig.\ref{FigMax}(a1). The agreement of the
topological phase transition point $\lambda $ and the sharp difference of
the quantum battery is very good for small loss as indicated by red. Hence,
the max[$W$] is a good topological-phase indicator.

\textbf{Charging power}: The charging power\cite{Binder,Camp,Sree} is
defined by the velocity of the ergotropy, that is 
\begin{equation}
P\left( t\right) \equiv \frac{dW\left( t\right) }{dt},
\end{equation}%
which measures how fast the quantum battery is charged. It is an important
marker of the quantum battery because quick charging is required in quantum
applications such as quantum amplifier of quantum internets or quantum error
corrections in quantum computing. We show the dynamics of the charging power
in Fig.\ref{FigDynamics}(a3) for the topological quantum battery and (b3)
for the second-order topological quantum battery. The maximum value of the
charging power is also an important marker of the quantum battery. We show
the max[$P$] in Fig.\ref{FigMax}(a3) and (b3). They also show sharp
difference between the topological and trivial phases.

\textbf{Discussion}: We have investigated a topological quantum battery
based on the SSH model and a second-order topological quantum battery based
on the breathing-Kagome lattice. In both systems, we introduce nonlinear
gain selectively at the output sites and uniform loss across all sites,
which have not been explored in previous studies\cite{Lu,Konar,Zhou}. These
realistic non-Hermitian ingredients stabilize persistent oscillations of the
quantum batteries. The maximum values of the ergotropy and the charging
power exhibit sharp topological phase transitions, demonstrating that the
topological edge and corner modes serve as highly efficient energy-storage
channels.

In the second-order topological quantum battery, an intriguing feature
emerges: two quantum batteries are charged coherently. Owing to the mirror
symmetry of the breathing-Kagome triangle, the two batteries become
phase-synchronized and acquire energy with the same phase, enabling coherent
collective charging.

Our results naturally generalize to a third-order topological quantum
battery realized in the breathing-Pyrochlore lattice\cite{EzawaKagome},
where three quantum batteries can be charged.

This work is supported by Grants-in-Aid for Scientific Research from MEXT
KAKENHI (Grant No. 23H00171).

\end{document}